\documentclass[sigconf,nonacm]{acmart}

\AtBeginDocument{%
  }

\renewcommand\footnotetextcopyrightpermission[1]{}

\begin{document}

\title{The Evolution of Digital Search:\\
From Blue Links to Delegated Decision-Making}

\author{
David M. Rothschild$^{*1}$ \quad
Nicole Immorlica$^{2,3}$ \quad
Brendan Lucier$^{2}$ \\[2pt]
Markus Mobius$^{2}$ \quad
Aleksandrs Slivkins$^{1}$
}

\affiliation{%
  \institution{$^{1}$Microsoft Research, New York, NY \quad
               $^{2}$Microsoft Research, Cambridge, MA \quad
               $^{3}$Yale University, New Haven, CT}
  \country{USA}}

\email{$^{*}$\,Corresponding Email: David@ResearchDMR.com}

\renewcommand{\shortauthors}{Rothschild et al.}

\begin{abstract}
Digital search is undergoing a fundamental transformation from a human-driven process of discovery to an agent-mediated system of delegated decision-making. In the traditional model of digital search, users translate intent into keyword-based queries, evaluate ranked lists of links, and execute decisions outside the search interface. 
In an AI-native world, users express goals in natural language, agents interpret these intentions, and outcomes are returned as recommendations or executed decisions. This shift moves search from a link-based user interface to an embedded system component, with implications for transparency, competition, and monetization.
The resulting system design problem raises key questions about information quality and access, trust, incentive alignment, and market structure. Early evidence from experimental agent-mediated marketplaces and economic theory suggests that small design choices, such as how stakeholders access information, how options are surfaced, and how actions are executed, have first-order effects on efficiency, competition, and the welfare of consumers and firms. We propose that the future of search will be determined not by incremental improvements in ranking algorithms and natural-language interfaces, but by the design of open, transparent, and competitive agentic systems that govern how decisions are made and how markets operate, highlighting a set of grand challenges at the intersection of AI, economics, and system design.
\end{abstract}


\maketitle

\section*{}

Digital search is undergoing a fundamental transformation. For three decades, the dominant system has remained remarkably stable: a person enters a keyword-based query, reviews a list of results, and decides what to do next. This \emph{legacy system} is now being replaced.

In an AI-native world, digital search is no longer the interface between a human user and the Internet. Instead, a new interface is emerging: delegated decision-making. Users increasingly rely on agents to interpret goals, evaluate options, and act on their behalf. Search, in this context, becomes a core component of the system that powers those decisions rather than an exposed tool that helps users discover links.

This transformation raises a central grand challenge: how to design search for agent-mediated systems that make decisions on behalf of users in ways that are efficient, trustworthy, and aligned with both user preferences and healthy market competition.

Throughout this paper, we define an \emph{agentic system} (in contrast to the \emph{legacy system}) as a structured ecosystem consisting of: users; consumer agents acting on behalf of users; discovery layers that intermediate and rank options; firms providing goods or services; and service agents acting on behalf of those firms \cite{rothschild2026agentic}. We use this definition as the baseline architecture for the emerging challenges we define and explore in this perspective.

We characterize this transition, and the challenges it raises, through four structural shifts: from ranked lists to delegated decisions, from an attention economy to the preference economy, from evaluating performance to verifying delegation, and from retrospective study to prospective research design. We conclude by listing the research challenges and the stakes.

\subsection*{From Ranked Lists to Delegated Decisions}

In the \emph{legacy system}, search is initiated and leveraged by humans. Users translate their intent into keywords, scan a visible list of results, and navigate links for information and tools that help achieve that intent~\cite{brin1998anatomy,kleinberg1999authoritative}. Firms compete for attention within this interface: whether by earning high placement organically or by paying for sponsored placement in search results~\cite{edelman2007internet,varian2007position}. Critically, the search engine’s role largely ends at discovery (despite the repeated efforts of search engines to incorporate more transactions). The actual decision---booking travel, purchasing a product, choosing a service---typically occurs off-platform.

This architecture shaped the modern Internet, placing the burden on users to translate intent through keyword-based queries and evaluate options themselves. As goals become more complex, this process is strained: keywords poorly capture nuanced preferences and comparing fragmented results requires substantial effort, creating pressure for systems that can better infer and act on user intent. An ecosystem for collecting and selling user data arose to help with intent prediction, albeit imperfectly.

The emerging AI-powered paradigm already looks fundamentally different. Users directly express intent in natural language, articulating goals such as 
\emph{find me a hotel similar to the one from last month but closer to downtown, optimizing for price and convenience over room size.}
An AI agent translates this intent into a sequence of actions. Rather than presenting a list of links, the system returns a synthesized recommendation or completes the task entirely.

In this shift, the ranked list ceases to be the primary interface between user and system. Instead, search operates beneath the surface as part of a broader process of delegated decision-making. While users may be able to access the ranked list (e.g., for audit or supporting information), they would typically choose not to and instead trust the AI to collect and synthesize the information itself.

\textit{\textbf{As users evolve from keyword-based queries and ranked lists of recommendations to natural language conversations delegating decisions, the choice set from search becomes hidden.}}

\subsection*{From Attention Economy to Preference Economy}

In an agentic system, users may never see most candidate options, encountering only curated recommendations or decisions. This shift transforms both the user experience and what determines outcomes in the system. The change is not primarily about who acts, but about what competition is organized around: from capturing attention to accurately representing and satisfying user preferences.

In the \emph{legacy system}, competition is centered on attention. Users face cognitive constraints, so visibility within the consideration set is decisive. Firms compete for clicks: ranking position determines attention, and attention drives outcomes. The unit of competition is therefore \emph{rank}; links are candidate objects, while rank allocates scarce exposure~\cite{athey2011position}.

In an \emph{agentic system}, preference inference becomes central. Consumer agents evaluate, compare, and filter options before a user observes them, drawing on richer, more structured models of user intent. These processes shape which options are considered and selected. Outputs may take the form of ranked recommendations, natural-language summaries, or direct transactions, while service agents representing firms increasingly respond and negotiate in real time with consumer agents. As with the \emph{legacy system}, firms could still signal relevance by paying to offset search costs, perhaps even by directly subsidizing agentic search~\cite{candelas2026subsidizing}. Even so, since decision-making is mediated by agents, the quality of preference modeling, how accurately systems infer and act on user intent, becomes the primary determinant of outcomes.

This shift gives rise to what can be understood as a \emph{preference economy}. In contrast to the legacy \emph{attention economy} where firms compete for user clicks, competition centers on the ability to represent and satisfy user preferences. This shift applies not only to agents that interpret and act on behalf of users or firms, but also to firms themselves: because consumer agents determine which options are considered in the first place, firms must compete to be chosen by agents, requiring that their offerings are interpretable, comparable, and closely aligned with inferred user preferences. Outcomes therefore depend not on attracting user attention, but on how effectively agents infer and act on user preferences \cite{rothschild2026agentic}.

\textit{\textbf{When users do not directly observe the choice set arising from search, competition depends on how accurately preferences are inferred, represented, and acted upon.}}

\subsection*{From Evaluating Performance to Verifying Delegation}

In the \emph{legacy system}, search and recommendation systems are evaluated using metrics suited to an attention economy: relevance, click-through rates, dwell time, and conversion. These metrics describe outcomes within the system rather than determine whether the system can function. Because users remain close to the underlying information---inspecting links, comparing sources, and exercising judgment---they can independently evaluate recommendations and correct mistakes. Measurement is therefore primarily retrospective: a way to assess performance after the fact.

This legacy structure implies a particular division of roles. Firms and platforms compete to influence user attention, often through adversarial strategies to improve rank, capture clicks, and shape engagement. No component of the system is required to act on the user's behalf: the user is the final decision-maker and their own advocate, providing a natural check on errors, manipulation, or low-quality outcomes. Alignment with the user is then a constraint, rather than a goal; a way for firms and platforms to entice user attention and participation.

In an \emph{agentic system}, that logic breaks down. If competition is now organized around user preferences, outcomes will depend on the extent to which agentic delegates are aligned with the preferences of the principal user~\cite{grossman1992analysis,holmstrom1979moral}.
For this to work, consumer agents must act on behalf of users: interpreting intent, evaluating tradeoffs, and sometimes executing decisions before users explicitly review the available options. Firms then compete on their ability to satisfy user preferences, with returns dependent on being selected by consumer agents operating within this environment.

This change alters the role of measurement itself. In an attention economy, evaluation metrics are primarily used to compare performance after users make their own decisions. In an agentic system, however, delegated decision-making requires users to trust that agents correctly represent and act on their preferences. Measurement therefore becomes an institutional necessity rather than merely an evaluative tool. If alignment cannot be observed, audited, or verified, delegation becomes difficult to sustain regardless of how accurately preferences are inferred.

Thus, a well-aligned consumer agent is not sufficient by itself: its performance depends on the system in which it operates, including which options are surfaced and how comparisons are structured. Alignment must therefore be achieved across multiple layers. Even though platforms and firms naturally act in their own interests, the delegation of user choices to an aligned consumer agent means that those interests will be best served when the system works with, rather than against, the user. This includes the responses of service agents and the behavior of the discovery layer that structures comparison, selection, and exposure.

Trust therefore shifts in character. Rather than relying primarily on direct inspection of results, users rely on agents to act on their behalf, and those agents depend on the surrounding system to surface options, enable comparisons, and support aligned decisions. While \emph{agentic systems} must provide mechanisms for auditing or inspecting underlying options, their adoption ultimately hinges on users’ willingness to delegate without full verification: if each decision requires inspection, the efficiency gains of delegation disappear.

\textit{\textbf{In an attention economy, measurement evaluates competition; in a preference economy, measurement enables delegation. Without mechanisms that make alignment observable, auditable, trustworthy, and resistant to manipulation, delegated decision-making cannot function at scale.}}

\subsection*{From Retrospective Study to Prospective Research Design}

Researching this transition requires empirical and theoretical tools suited to systems that are still emerging. The challenge is not simply to study a mature system, but to predict and steer a disruption before its dominant technologies, norms, and market structures have fully stabilized. \emph{Agentic systems} are dynamic: agents interact with one another, adapt to incentives, and make decisions on behalf of users. While forward-looking research is always desirable, it is especially true under these conditions 
that effective research cannot be purely backward-looking or static.

The transition to agent-mediated search reopens a rich design space for platforms and discovery. Economic models may diverge: advertising may persist but become embedded within agent decision processes; subscription-based systems, vertically integrated platforms, open or decentralized systems, or hybrid approaches may become more viable. As agents intermediate between users and providers, the structure of incentives across users, platforms, and firms becomes potentially contestable; that is, open to entry and competing designs, potentially reducing centralized control. All of these possibilities require novel research to understand the potential evolution of future systems.

Recent work has begun to explore this space through synthetic \emph{agent-mediated marketplaces}. In these environments, both consumers and firms are represented by autonomous agents, and market outcomes emerge from their interactions. Experimental platforms such as our Magentic Markets platform \cite{bansal2025magentic} allow researchers to evaluate how design choices affect behavior and welfare.

Early results highlight several consistent patterns, although the extent to which these reflect structural dynamics versus model- or prompt-specific artifacts remains an open question. For example, we have published on how agents exhibit strong biases toward early offers, creating advantages for speed over quality; systems that perform well under controlled conditions degrade as complexity increases, suggesting limits to local optimization in multi-agent environments; and small changes in how options are surfaced or ranked to agents can significantly alter outcomes, indicating that platform design shapes which market equilibria emerge.

Complementing these experimental approaches, theoretical work shows that without adequate information flow, agentic search can (in some scenarios) degrade market outcomes \cite{lucier2026agentic}. Allowing agents to consider hundreds or thousands of potential matches, rather than the few a human might inspect directly, is not sufficient if the system cannot learn which attributes or interactions produce high-quality matches. Without adequate information flow, high-quality options may lose visibility, rankings may deteriorate, and outcomes can become inefficient despite greater nominal access to information. Moreover, the apparent quality of agentic recommendations can suppress a human user's appetite for exploring different options, cutting off a natural release valve for poor information aggregation at the market level. This illustrates a broader challenge: \emph{agentic systems} requires not only more information, but mechanisms that make information interpretable, shareable, and usable within the system while preserving privacy and ensuring sufficient exploration to learn which attributes and interactions produce high-quality matches.

These early studies also suggest that existing ranking and evaluation frameworks, which were largely designed for human-facing search, are insufficient as a foundation for the study of agentic systems. A central direction for future work is the development of agent-native infrastructure: indexes that make agents discoverable and verifiable, mechanisms that generate reliable signals of quality, and evaluation systems that sustain trust, alignment, and competition over time. The key question is under what conditions \emph{agentic systems} produce stable, competitive, and welfare-enhancing outcomes.

\textit{\textbf{There is an immediate need for research on agentic delegation as a system design framework, before agentic markets harden into durable institutions.}}

\subsection*{Research Challenges}

This all gives rise to a set of interrelated research and development challenges. These directions are conceptually distinct but mutually reinforcing, reflecting the fact that in agentic systems, information generation, alignment, and market design are tightly coupled. More broadly, these challenges trace the transition described throughout the paper: search becomes less visible to users; attention gives way to preference as the basis of competition; and, in turn, stakeholders must align around representing and acting on those preferences. This shift creates new opportunities for economic systems, but also raises fundamental questions about how to guide this transition toward efficient, competitive, and trustworthy outcomes.

\begin{itemize}
    \item \textbf{Information} - Agent Information Generation and Filtering: If agents become the primary intermediaries through which information is generated, selected, and aggregated for users, what biases, omissions, or failures arise in how information is constructed and filtered?
    
    \item \textbf{Representation} - Agent Trust, Alignment, and Accountability: Given that agents act on behalf of users, how can agentic search systems ensure that decisions faithfully reflect user preferences while remaining interpretable, auditable, and robust to manipulation?

    \item \textbf{Governance} - Market Design and Mechanism Structure: How should system-level rules, such as ranking, pricing, and participation rules, be designed when agents --- rather than humans --- are the primary actors?  What monetization strategies will platforms employ, and how will efficiency gains from agentic delegation be divided among stakeholders?

    \item \textbf{Dynamics} - Market Learning and Feedback: How do agentic systems most effectively learn from interactions with agents over time?  How do feedback loops shape market outcomes, stability, and welfare in complex, adaptive environments?

    \item \textbf{Response} - Strategic Behavior and Incentives: How will firms, platforms, and agents adapt to agent-mediated environments, and what new forms of manipulation, optimization, and collusion will emerge?
\end{itemize}

These are not purely technical questions. Indeed, the largest disruptions arising from agentic systems are the disruptions to workflow and incentives~\cite{rothschild2026augmentation}, which are poised to reshape how users conceptualize and interface with information and search. These questions will determine who captures value, how markets function, and whether agentic systems expand or constrain economic opportunity.

\subsection*{The Stakes}

Search will not disappear. But it will no longer be the primary interface through which users engage with information and, as a result, it will no longer serve as the central organizing paradigm of the Internet. Instead, it becomes embedded within a broader system of delegation where users articulate goals, agents act on their behalf, and markets operate at machine speed. Search becomes one component of a larger decision-making infrastructure, rather than the interface through which users directly engage with information.

This shift changes the core research question. It is no longer simply how information is retrieved, but how decisions are made: which options are generated, how they are filtered and compared, who determines the criteria for evaluation, and under what incentives these processes operate.

The \emph{legacy system} was not inevitable. It reflected a set of design choices, most notably the dominance of advertising-funded models that aligned incentives around attention and engagement. Combined with centralized data accumulation and processing, and communication frictions that induced lock-in, these choices produced concentrated control over discovery and attention in a small number of platforms.

There is the potential for \emph{agentic systems} to disrupt a core monetization channel of the \emph{legacy system}, which ran through search advertising. As agents intermediate between users and firms, value no longer primarily flows through exposure, but through the mechanisms that govern decision-making. Economic activity therefore shifts toward influencing the design of these mechanisms, even as those designs are shaped by the incentives and competitive pressures they create.

The research and design of \emph{agentic systems} therefore carries significant economic and societal consequences. The mechanisms governing how agents access information, evaluate options, and execute actions shape market efficiency, the structure of competition, the pace of innovation, and the extent to which users trust outcomes. In a delegated system, these mechanisms determine how value is created, allocated, and sustained.

The risks follow directly from these design choices. If agents optimize for speed or engagement, markets may reward early or strategically positioned responses rather than better ones. If decision processes are opaque, users cannot assess whether outcomes reflect their preferences, and trust may erode. If platforms tightly control interfaces and standards, competition may narrow; if systems are open, modular, and transparent, they may instead foster entry, innovation, and user control.

What distinguishes this moment is not only that design choices are reopened, but that their consequences are less visible and more persistent. As decision-making moves under the hood, outcomes become harder to inspect, correct, or contest.

The next era of search will not be defined by better answers, but by better systems. The choices made now---about information flows, representation, governance, dynamics, and incentives---will determine whether \emph{agentic systems} concentrate power and constrain what can be known, chosen, and achieved, or whether they expand opportunity, sustain competition, and build trust.

\begin{acks}
We thank Patrick Jordan for many helpful conversations.
\end{acks}

\section*{Ethics and Privacy Statement}

This paper explores the transition from human-directed search to agent-mediated decision systems and highlights the central role of system design in shaping outcomes. Because these systems increasingly act on behalf of users, they raise important ethical and privacy considerations. Agent-mediated search concentrates decision-making authority within opaque systems, creating risks related to bias, manipulation, and misalignment with user preferences. In addition, such systems may require access to sensitive user data in order to accurately represent preferences, raising concerns around data use, security, and consent. These risks are not incidental, but emerge directly from design choices. Mechanisms governing information access, ranking, and action execution can either amplify or mitigate harms. We therefore argue that transparency, auditability, and alignment with user interests should be treated as core design constraints in agentic systems, alongside efficiency and performance.

\bibliographystyle{plainnat}
\bibliography{sample-base}

\end{document}